# On the Production of Bright RR Lyræ Variables in Metal-Rich Globular Clusters

A. V. Sweigart and M. Catelan

NASA/Goddard Space Flight Center
Laboratory for Astronomy and Solar Physics, Code 681
Greenbelt, MD 20771, USA

e-mail: sweigart@bach.gsfc.nasa.gov; catelan@stars.gsfc.nasa.gov

## ABSTRACT

Recent HST observations of the metal-rich bulge clusters NGC 6388 and NGC 6441 by Rich et al. (1997) have found that the horizontal branches (HBs) in these clusters slope upward with decreasing *B–V*. Such an upward slope in the HB morphology is not predicted by canonical HB models. Moreover, it cannot be produced by either a greater cluster age or enhanced mass loss along the red-giant branch (RGB). The peculiar HB morphology in these clusters may provide an important clue for understanding the second-parameter effect.

We have carried out extensive evolutionary calculations and numerical simulations in order to understand the cause of the sloped HBs in NGC 6388 and NGC 6441. Three scenarios have been investigated:

- A high cluster helium abundance scenario, where the HB morphology is determined by long blue loops;
- A rotation scenario, where the core mass in the HB models is increased by internal rotation during the RGB phase;
- A helium-mixing scenario, where deep mixing on the RGB enhances the envelope helium abundance.

All three of these scenarios predict sloped HBs with anomalously bright RR Lyræ variables. We compare this prediction with the properties of the two known RR Lyræ variables in NGC 6388 as well as with the properties of the metal-rich field RR Lyræ variables and V9 in 47 Tuc.

# 1. Introduction

The Galactic bulge globular clusters play a fundamental rôle in determining the formation history of our Galaxy (e.g., Renzini 1994; Ortolani et al. 1995; Cravet et al. 1997). Moreover, they provide a metal-rich template for interpreting the integrated spectra of the old stellar populations in elliptical galaxies. For these reasons the properties of the bulge clusters (their ages, composition) are crucial for understanding both Galactic evolution and the evolution of old metal-rich stars.

Recent HST observations of the metal-rich bulge clusters NGC 6388 and NGC 6441 by Rich et al. (1997) have revealed two intriguing properties (see **Figure 1**). First, these clusters contain a significant population of hot HB stars and therefore exhibit the well-known second-parameter effect. Second, the horizontal branches (HBs) of these clusters have a pronounced upward slope with decreasing $B\!-\!V$, with the mean HB luminosity at the top of the blue tail being nearly 0.5 mag brighter in $V$ than the well-populated red clump. A similar upward slope is clearly present in the red clump itself (Piotto et al. 1997).

The second parameter effect has often been attributed to differences in age or mass loss on the red-giant branch (RGB). However, such differences cannot be reconciled with the upward sloping HBs in NGC 6388 and NGC 6441. We emphasize therefore that these clusters may be providing an important clue for understanding the origin of the second parameter effect.

In this paper we investigate the following three scenarios for explaining the hot HB population and, in particular, the sloped HBs observed in NGC 6388 and NGC 6441:

High $Y$ scenario:

> high cluster helium abundance $Y \Rightarrow$ unusually long blue loops during the HB evolution

Rotation scenario:

> rotation $\Rightarrow$ increase in core mass, making HB both bluer and brighter

Helium-mixing scenario:

> deep mixing along the RGB $\Rightarrow$ increase in the HB envelope helium abundance

These scenarios predict sloped HBs with anomalously bright RR Lyræ variables. We test this prediction by using the two known RR Lyræ variables in NGC 6388 and the variable V9 in 47 Tuc.

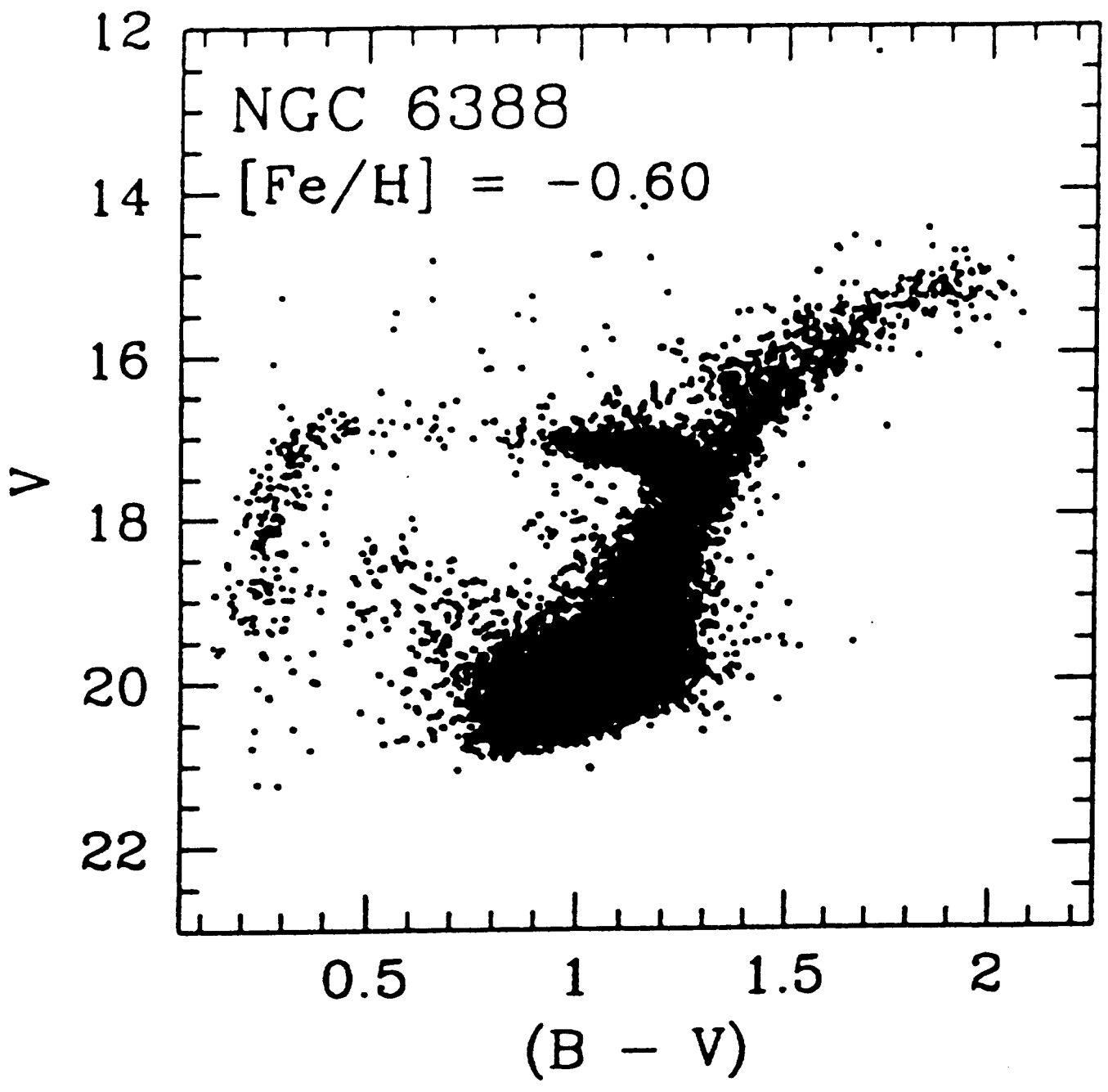

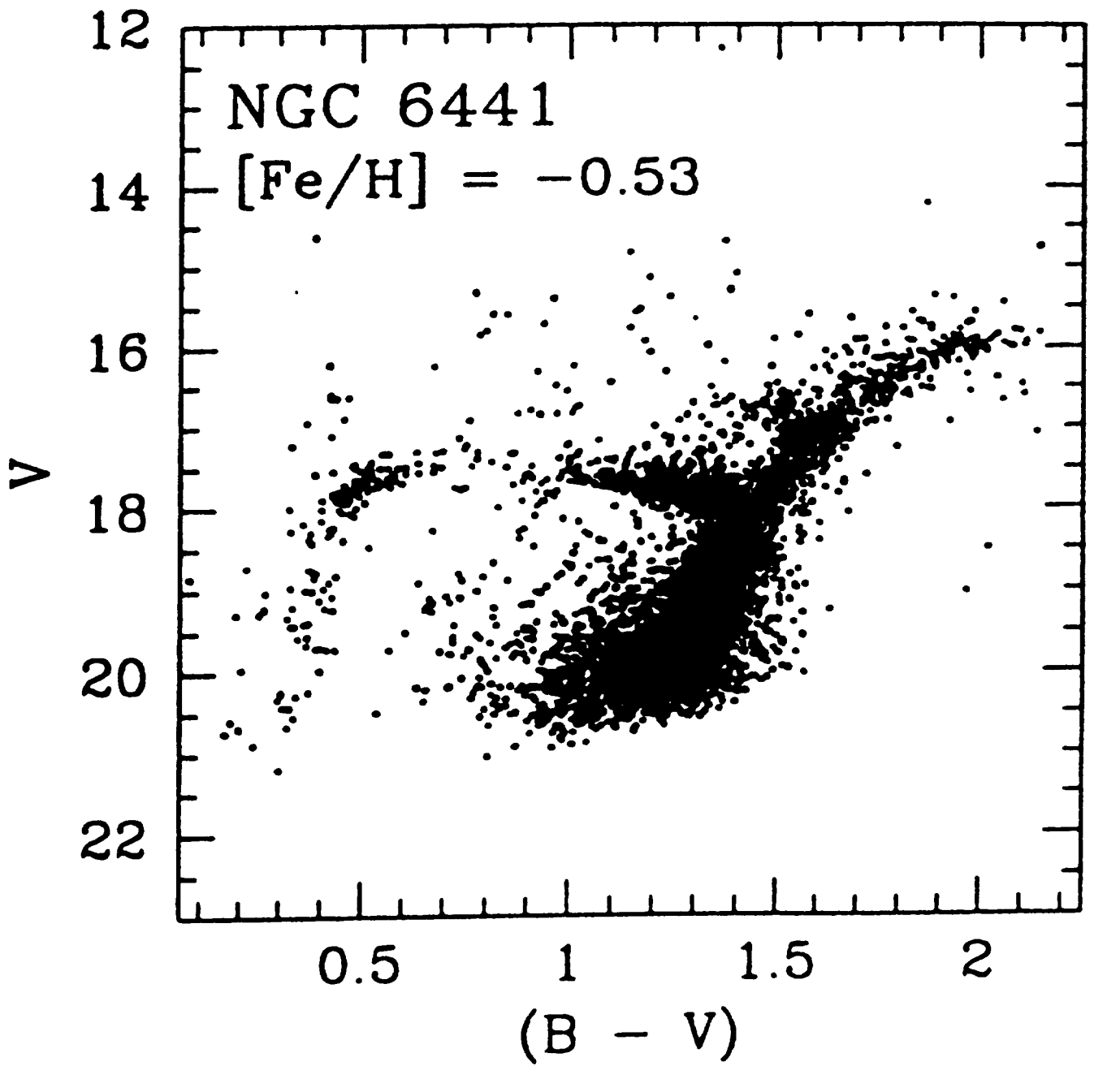

**Figure 1.** Color-magnitude diagrams (CMDs) for the metal-rich bulge globular clusters NGC 6388 and NGC 6441 taken from Fig. 1 of Rich et al. (1997). Note the upward slope of the HB with decreasing *B*–*V* and the suggestion of HB bimodality.

# 2. Canonical HB Morphology

In order to explore the HB morphology predicted by canonical models for NGC 6388 and NGC 6441, we have constructed a grid of HB sequences with a main-sequence helium abundance $Y_{MS}$ = 0.23 and a scaled-solar heavy-element abundance $Z$ = 0.006 (i.e., [Fe/H] = –0.5). Using these sequences together with the Kurucz (1992) color-temperature transformations and bolometric corrections, we then carried out an HB simulation with the mean mass and mass dispersion chosen to give a well-populated red and blue HB. This simulation, shown in **Figure 2**, demonstrates that the HB morphology predicted by canonical models is completely flat unlike the observed HBs in NGC 6388 and NGC 6441.

Increasing the assumed age or RGB mass loss would move a red HB star blueward along the distribution in Figure 2 but would not increase its luminosity. This indicates that the blue HB population in NGC 6388 and NGC 6441 does not arise from either an older cluster age or greater RGB mass loss.

We have also investigated whether the slope of the red HB in NGC 6388 and NGC 6441 could be due to differential reddening. **Figure 3** shows an HB simulation for the red HB which includes a uniform differential reddening $\Delta E(B{-}V)$ of 0.10 mag, a value which considerably exceeds the amount estimated for NGC 6388 and NGC 6441 by Piotto et al. (1997). We see that even this large amount of differential reddening cannot produce a sloped red HB. Thus differential reddening can be excluded as the cause of the sloped red HBs in these clusters (see also Catelan & de Freitas Pacheco 1996).

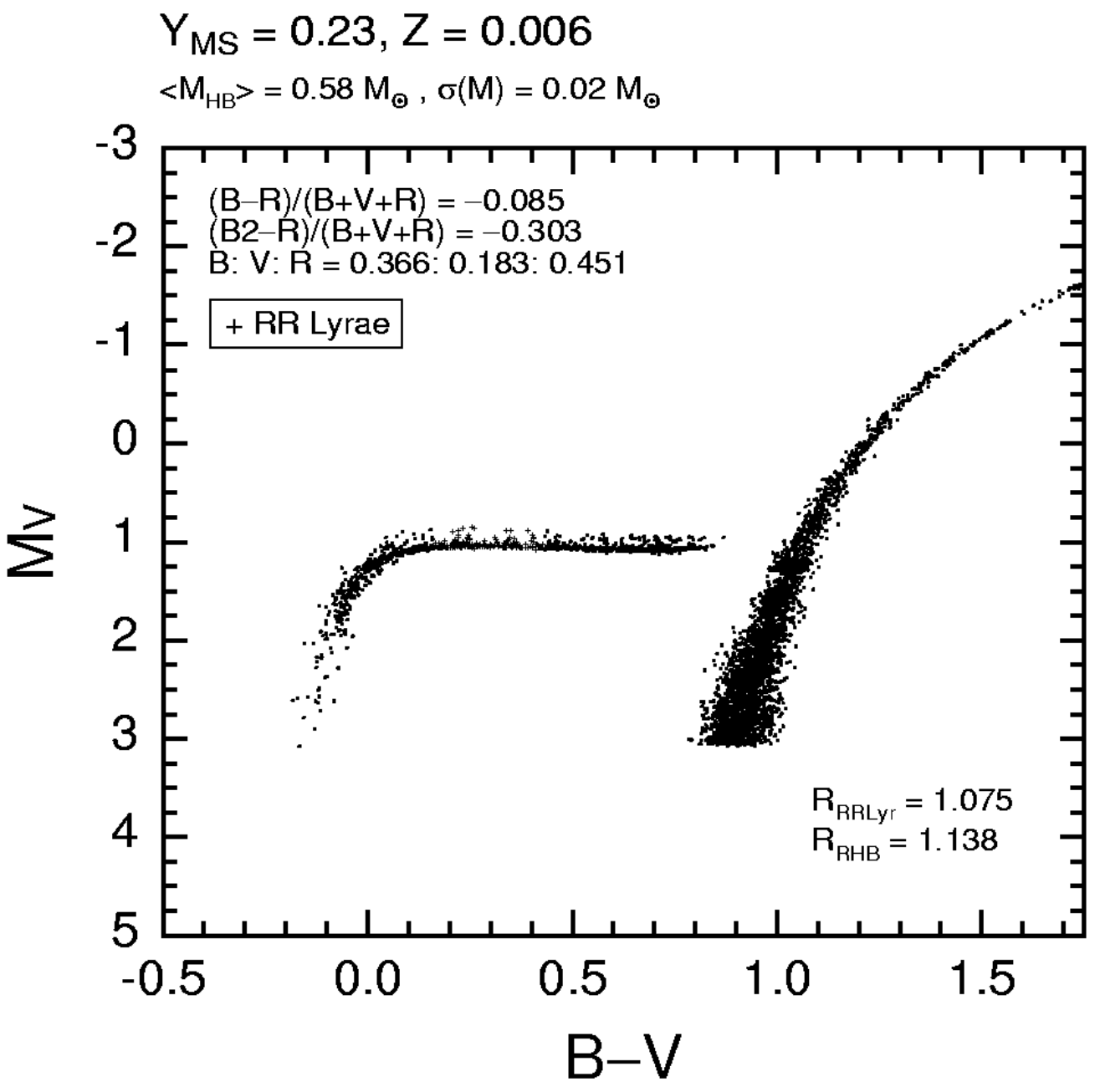

**Figure 2.** Canonical simulation of the HB morphology for a helium abundance $Y_{\rm MS}$ = 0.23 and a heavy-element abundance $Z$ of 0.006. Note that the HB is completely flat between $B\!-\!V \approx 0.1 - 0.9$

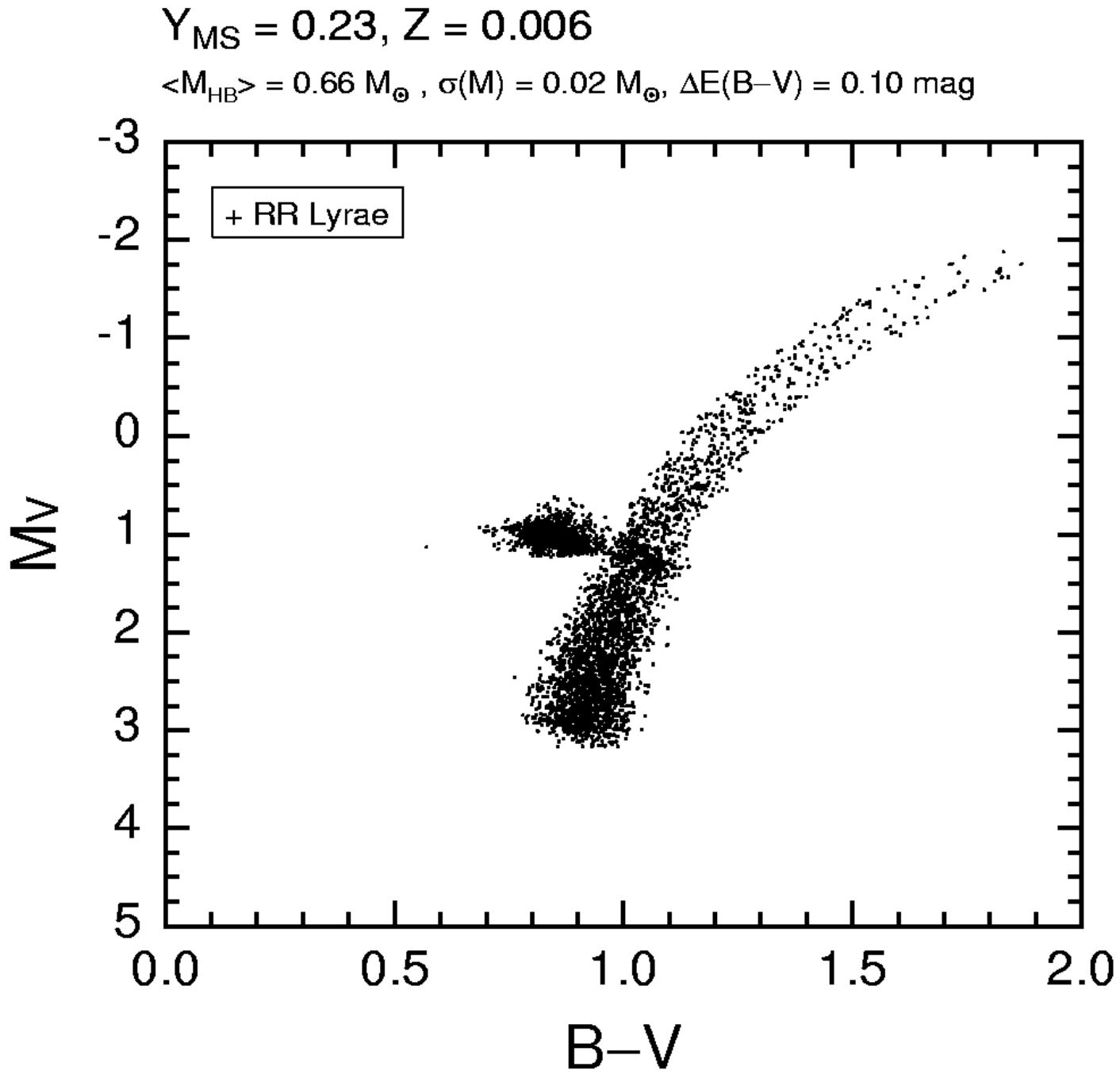

**Figure 3.** HB simulation for the red HB population in NGC 6388 and NGC 6441 including a differential reddening $\Delta E(B\!-\!V)$ of 0.10 mag. Note that the slope of the red HB remains quite flat

# 3. High *Y* Scenario

Red HB stars evolve along blue loops during most of their HB lifetime. Normally these loops cover only a small range in effective temperature or *B–V*. For larger helium abundances, however, these blue loops can become considerably longer, reaching higher effective temperatures and deviating more in luminosity from the zero-age HB (ZAHB). Thus, at least qualitatively, one would expect the HB for a sufficiently high helium abundance to show the sloped structure evident in NGC 6388 and NGC 6441 (Catelan & de Freitas Pacheco 1996).

To test this possibility, we have constructed a grid of HB sequences for $Y_{MS}$ = 0.23, 0.28, 0.33, 0.38 and 0.43 and $Z$ = 0.006. Similar grids for $Z$ = 0.002 and 0.01716 (= $Z_\odot$) were also constructed to explore any dependence on metallicity. For each ($Y_{MS}$, $Z$) combination we first evolved a star up the RGB without mass loss and then through the helium flash to obtain a ZAHB model at the red end of the HB. Lower mass ZAHB models were then obtained by removing mass from the envelope of this high mass ZAHB model. In total, 357 HB sequences were computed. As expected, the HB simulations for $Y_{MS}$ = 0.23, 0.28 and 0.33 did not show a significant slope. However, when $Y_{MS}$ was increased to 0.38 and 0.43, a pronounced HB slope was found, as illustrated in **Figures 4a,b**. The HB slope for $Y_{MS}$ = 0.43 in Figure 4b is close to that in NGC 6388 and NGC 6441. There is also a hint of bimodality.

In these simulations we did not try to reproduce the ratio of blue and red HB stars in NGC 6388 and NGC 6441. This ratio can be adjusted by varying the mean mass and mass dispersion in the HB simulations. We note that the HBs in Figures 4a,b are quite bright, implying a large pulsation period for the RR Lyræ variables. Furthermore, these simulations predict a large value for the *R* ratio (= HB/RGB stars), which can be tested once number counts are available for these clusters.

Conclusion: For sufficiently high helium abundances the HB should slope upward with decreasing *B–V*.

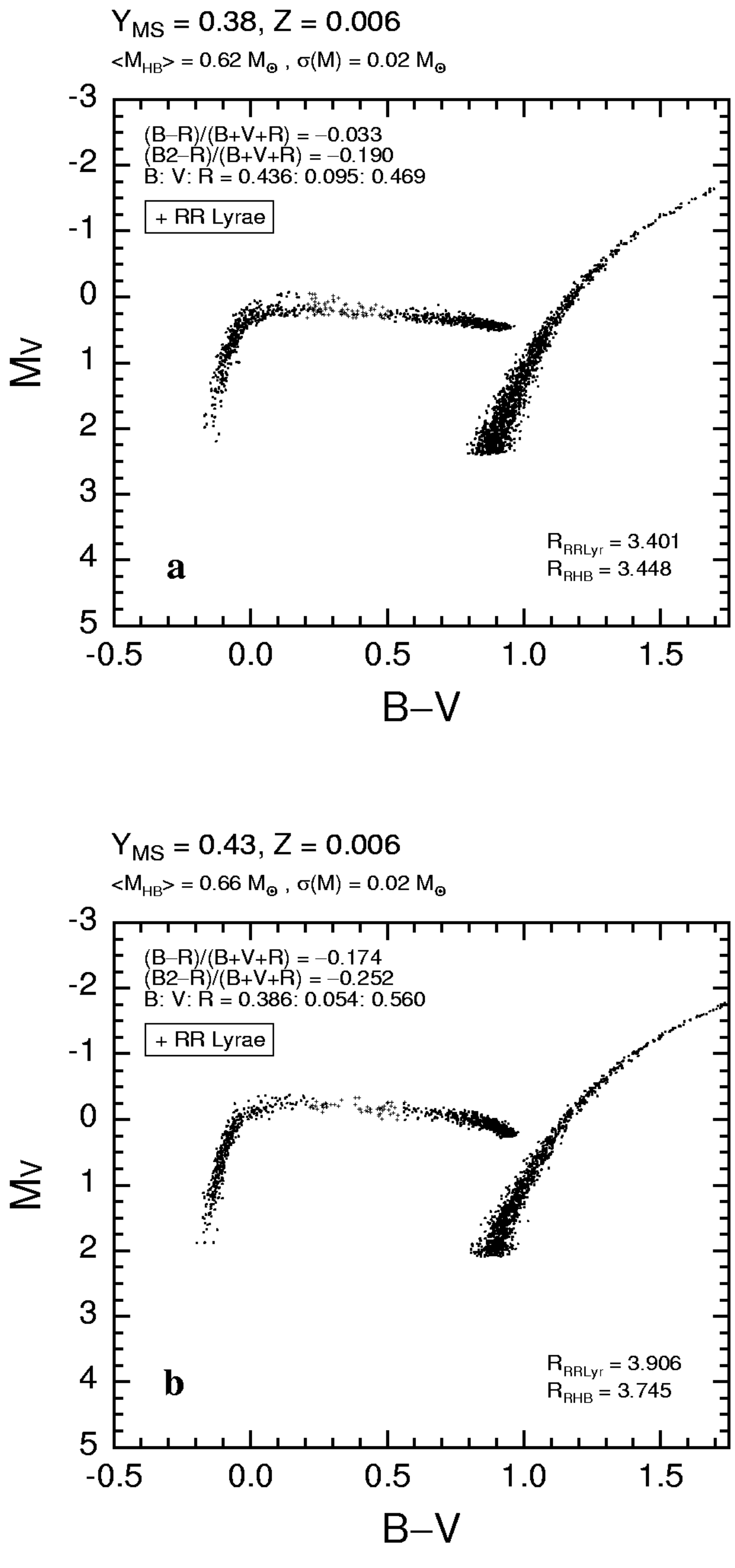

**Figure 4.** HB simulations for high helium abundances: $Y_{MS}$ = 0.38 (panel a) and $Y_{MS}$ = 0.43 (panel b). Note the upward slope of the HB with decreasing *B–V*.

# 4. Rotation Scenario

Since the work of Mengel & Gross (1976) it has been known that rotation during the RGB phase can delay the helium flash. As reviewed by Renzini (1977), this would have two consequences for the subsequent HB evolution. First, it would increase the helium-core mass and hence the HB luminosity. Second, it would lead to enhanced mass loss near the tip of the RGB and thus to a smaller HB envelope mass. The net effect would be a shift in the HB location towards higher effective temperatures and luminosities. Thus, at least qualitatively, one might also expect a range in rotation to produce an upward sloping HB morphology. Could this be the explanation for the sloped HBs in NGC 6388 and NGC 6441?

To answer this question, one needs to know how much extra mass is lost at the tip of the RGB when the core mass exceeds its canonical value. We determined this by evolving a number of sequences up the RGB for various values of the Reimers mass loss parameter $\eta_R$. In these sequences we turned off all helium burning, thereby permitting the core mass to exceed its canonical value without igniting the helium flash. Using these sequences, we were able to determine how the final mass $M$ decreases with increasing core mass $M_c$ for each value of $\eta_R$. The $M$–$M_c$ relations defined in this manner were then used to compute grids of HB sequences for $\eta_R$ = 0.1, 0.2, 0.3 and 0.4.

Two simulations for this rotation scenario are presented in **Figures 5a,b** for $\eta_R$ = 0.2 (panel a) and 0.4 (panel b). The mean mass $\langle M_{HB} \rangle$ for the Gaussian mass distribution in each simulation was taken to be the mass of a canonical model without rotation. Each simulation covers the mass range $M < \langle M_{HB} \rangle$. In both simulations in Figures 5a,b the HB slopes upward, especially in the $\eta_R$ = 0.2 case. We are currently carrying out further simulations to determine how this result depends on the assumed cluster age and helium abundance.

Conclusion: The increase in the core mass $M_c$ and decrease in the total mass $M$ caused by rotation along the HB may produce a sloped HB morphology.

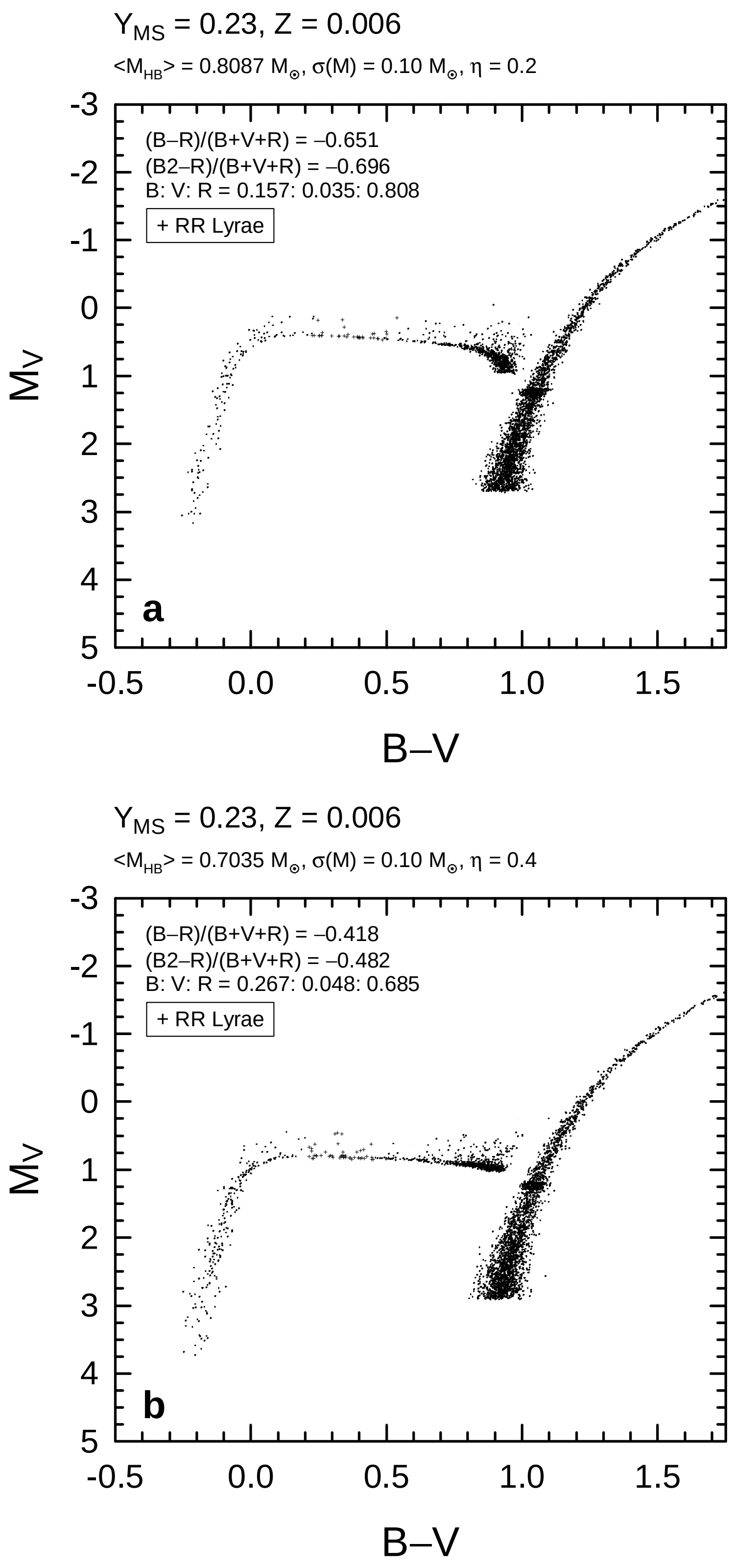

**Figure 5.** HB simulations including the effects of rotation along the RGB for the Reimers mass loss parameter $\eta_R$ = 0.2 (panel a) and 0.4 (panel b). Note the upward slope of the HB with decreasing *B*–*V*.

# 5. Helium-Mixing Scenario

The observed abundance variations in globular cluster red-giant stars indicate that these stars are able to mix nuclearly processed material from the vicinity of the hydrogen shell out to the surface (Kraft 1994). The observed variations in the aluminum abundance are particularly important because they require the mixing to penetrate into the hydrogen shell. Thus any mixing process which dredges up aluminum will also dredge up helium. Besides increasing the envelope helium abundance such mixing would also increase the RGB tip luminosity and thus the amount of mass loss. Consequentely a red-giant star which undergoes such helium mixing will arrive on the HB with both a higher envelope helium abundance and a lower mass and hence will be both hotter and brighter than the corresponding canonical star (Sweigart 1997a,b). This suggests, at least qualitatively, that helium mixing might also lead to a sloped HB morphology.

To investigate this possibility, we have computed a set of 99 sequences for various amounts of helium mixing. Each sequence was evolved up the RGB and then through the helium flash and HB phases. Two cases were considered, namely, ($M$, $Y_{MS}$, $Z$) = (0.95, 0.23, 0.006) and (0.87, 0.28, 0.006), where the mass in each case was chosen to give an age of 13 Gyr at the tip of the RGB. The values of $\eta_R$ used in these sequences were 0.4, 0.5 and 0.6.

Two simulations based on these helium-mixed sequences are presented in **Figures 6a,b**. The red end of the HB in these simulations is populated by models which did not undergo helium mixing, i.e., by canonical models. The masses of these canonical models were used as the mean mass $\langle M_{HB} \rangle$ for defining the Gaussian mass distribution, as indicated at the top of each panel in these figures. Each simulation covers the mass range $M < \langle M_{HB} \rangle$.

As the helium mixing increases, the HB tracks shift blueward, giving rise to the pronounced upward slope evident in Figures 6a,b. In fact, the HB slope in Figure 6a slightly exceeds the observed slope in NGC 6388 and NGC 6441. There is, moreover, a hint of bimodality in this simulation.

Conclusion: Helium mixing during the RGB phase can lead to a sloped HB morphology.

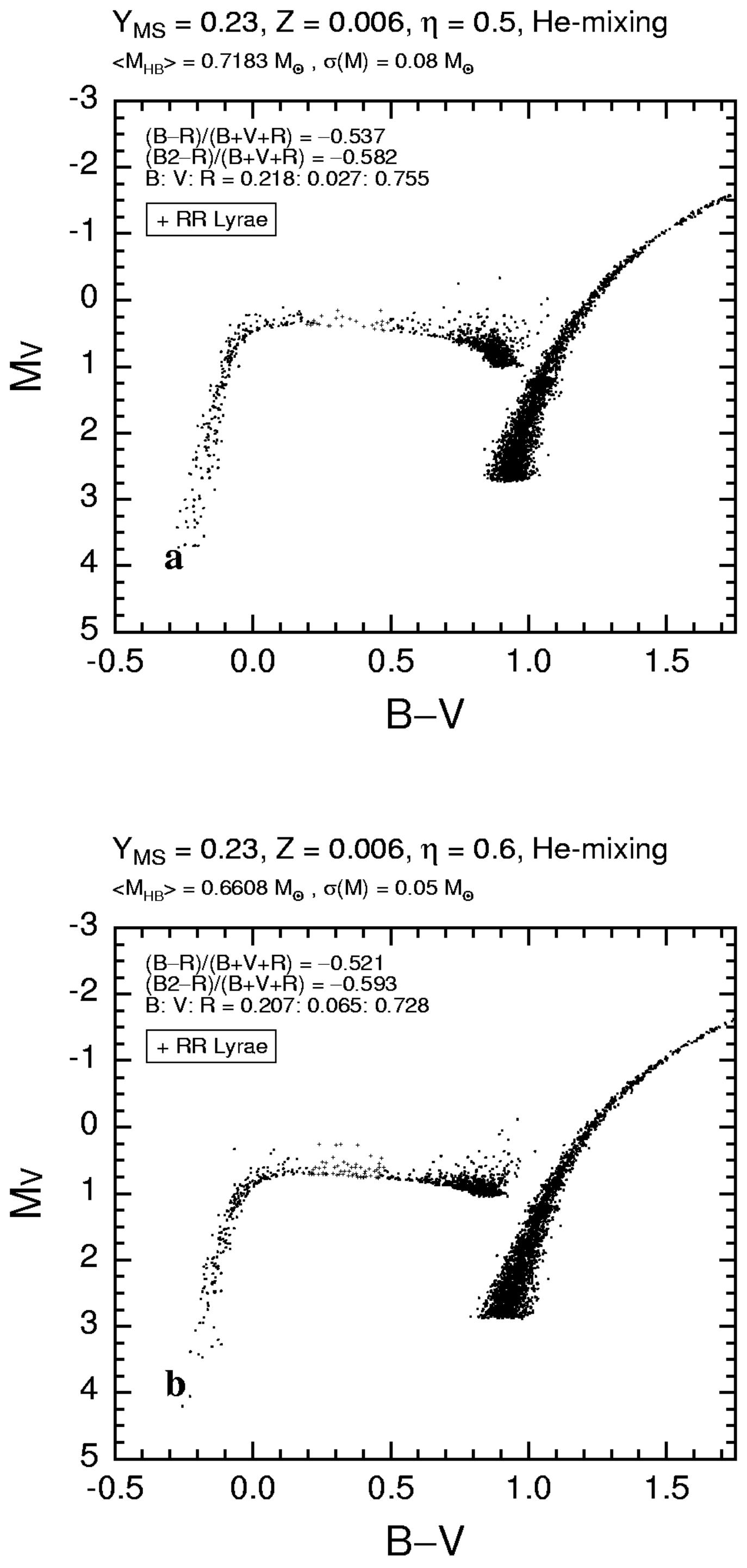

**Figure 6.** HB simulations including helium mixing along the RGB for the Reimers mass loss parameter $\eta_R$ = 0.5 (panel a) and 0.6 (panel b). Note the upward slope of the HB with decreasing *B–V*.

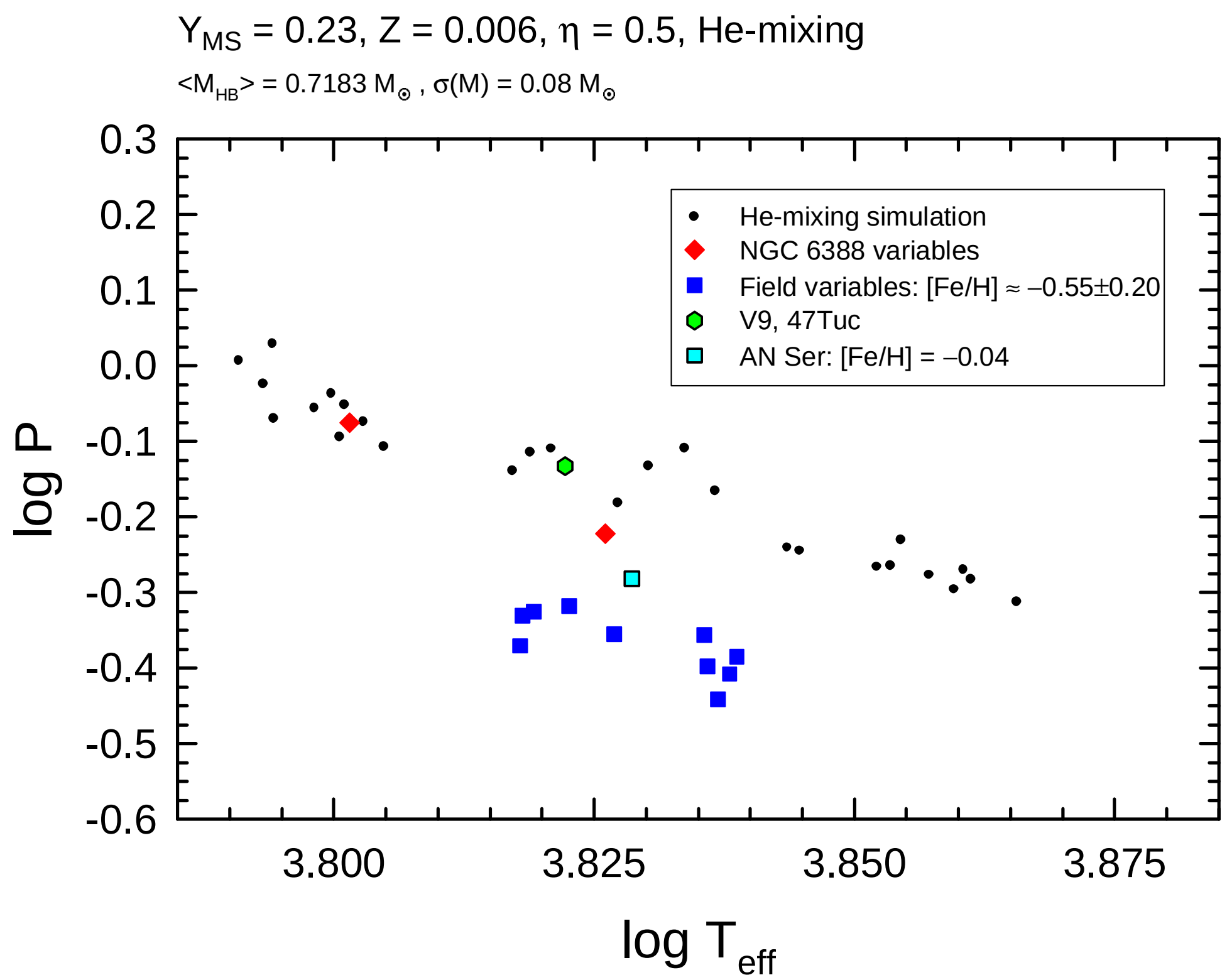

**Figure 7.** Period-temperature diagram for the RR Lyræ variables in NGC 6388 (◆). For comparison, the positions of field variables of comparable metallicity (■), as well as those of V9 in 47 Tuc (◆) and the very metal-rich, long-period field RR Lyræ variable AN Ser (■) are also given. Model predictions are shown as dots.

## 6. Constraints from RR Lyræ Variables

We compared the predictions of the above models with the observed period-temperature diagram for the two RRab Lyræ variables in NGC 6388 (Silbermann et al. 1994). Temperatures were determined as in Catelan et al. (1997). The result of one of these comparisons is given in **Figure 7**, where local field variables of metallicity similar to that of NGC 6388 and 47 Tuc are also displayed. The [Fe/H] values for the field variables were taken from Layden (1994); amplitudes and periods were obtained primarily from Blanco (1992), but in some cases also from Sandage (1990). Blazhko variables were avoided.

As one can clearly see, the variable stars in the clusters NGC 6388 and 47 Tuc seem substantially brighter than most field stars of comparable metallicity. The cluster variables seem in good agreement with the expected location from the displayed models, which in the specific case shown (corresponding to Figure 6a) have $Y_{HB} \sim 0.35$ in the instability strip region.

## References

Blanco, V. M. 1992, AJ, **104**, 734
Catelan, M., & de Freitas Pacheco, J. A. 1996, PASP, **108**, 166
Catelan, M., Sweigart, A. V., & Borissova, J. 1997, poster paper at this conference
Cravet, F. L., Guarnieri, M. D., Renzini, A., & Ortolani, S. 1997, in Advances in Stellar Evolution, ed. R. T. Rood & A. Renzini (Cambridge: Cambridge University Press), 59
Kraft, R. P. 1994, PASP, **106**, 553
Kurucz, R. L. 1992, in The Stellar Populations of Galaxies (IAU Symp. 149), ed. B. Barbuy & A. Renzini (Dordrecht: Kluwer), 225
Layden, A. C. 1994, AJ, **108**, 1016
Mengel, J. G., & Gross, P. G. 1976, Ap&SS, **41**, 407
Ortolani, S., Renzini, A., Gilmozzi, R., Marconi, G., Barbuy, B., Bica, E., & Rich, R. M. 1995, Nat, **377**, 701
Piotto, G., et al. 1997, in Advances in Stellar Evolution, ed. R. T. Rood & A. Renzini (Cambridge: Cambridge University Press), 84
Renzini, A. 1977, in Advanced Stages of Stellar Evolution, ed. P. Bouvier & A. Maeder (Geneva: Geneva Obs.), 149
Renzini, A. 1994, A&A, **285**, L5
Rich, R. M. et al. 1997, ApJ, **484**, L25
Sandage, A. 1990, ApJ, **350**, 631
Silbermann, N. A., Smith, H. A., Bolte, M., & Hazen, M. L. 1994, AJ, **107**, 1764
Sweigart, A. V. 1997a, ApJ, **474**, L23
Sweigart, A. V. 1997b, in The Third Conference on Faint Blue Stars, ed. A. G. D. Philip, J. Liebert & R. A. Saffer (Schenectady: L. Davis Press), in press (astro-ph/9708164)